\documentclass{PoS}

\usepackage{floatrow}
\usepackage{slashed}

\newcommand{\ahvp}{a_\mu^{\rm hvp}}

\newcommand{\fm}{{\rm{fm}}}

\newcommand{\MeV}{\mathrm{MeV}}

\title{The leading hadronic vacuum polarization contribution to the muon anomalous magnetic moment using $N_f=2+1$ O($a$) improved Wilson quarks}

\ShortTitle{The leading hadronic vacuum polarization contribution to the muon $(g-2)$}

\author{\speaker{Antoine G\'erardin}$^{1}$,
  Marco C\`e$^{3}$, Georg von Hippel$^{2}$, Ben H\"orz$^{4}$,
  Harvey Meyer$^{2,3}$, Daniel Mohler$^{2,3}$, Konstantin Ottnad$^{2}$, Jonas Wilhelm$^{2}$,
  Hartmut Wittig$^{\,2,3}$ \\ 
  	\llap{$^1$}John von Neumann Institute for Computing, DESY, Platanenallee 6, D-15738 Zeuthen, Germany\\
        \llap{$^2$}PRISMA$^+$ Cluster of Excellence and Institute for Nuclear Physics, University of Mainz, D-55099 Mainz, Germany\\
        \llap{$^3$}Helmholtz Institute Mainz, University of Mainz, D-55099 Mainz, Germany\\
        \llap{$^4$}Nuclear Science Division, Lawrence Berkeley National Laboratory, Berkeley, CA 94720, USA\\
        E-mail:
        \email{antoine.gerardin@desy.de}}

\abstract{We present a lattice calculation of the leading hadronic contribution to the anomalous magnetic moment of the muon. This work is based on a subset of the CLS ensembles with $N_f = 2+1$ dynamical quarks and a quenched charm quark. Noise reduction techniques are used to improve significantly the statistical precision of the dominant light quark contribution. The main source of systematic error comes from finite size effects which are estimated using the formalism described in Ref.~\cite{Bernecker:2011gh} and based on our knowledge of the timelike pion form factor. 
The strange and charm quark contributions are under control and an estimate of the quark-disconnected contribution is included. Isospin breaking effects will be studied in a future publication but are included in the systematic error using an estimate based on published lattice results. Our final result, $\ahvp = (720.0\pm 12.4 \pm 6.8)\times 10^{-10}$, has a precision of about 2\%.
\vspace{0.5cm}
\begin{flushright}
  MITP/19-075\\
  DESY 19-196
\end{flushright} }

\FullConference{37th International Symposium on Lattice Field Theory - Lattice2019\\
		16-22 June 2019\\
		Wuhan, China}

\begin{document}

%---------------------------------------------------------------------------------------------------------
\section{Introduction}
%---------------------------------------------------------------------------------------------------------

The anomalous magnetic moment of the muon is a promising observable for searches of new physics. It has been measured with a precision of 0.5 ppm by the Brookhaven experiment~\cite{Bennett:2006fi} and computed to a similar precision within the Standard Model, while a tension of about 3.4 standard deviations is observed. Two new experiments at Fermilab and J-PARC~\cite{Grange:2015fou,Mibe:2011zz} plan to reduce the error by a factor 4 and a similar reduction of the theory error is highly desired. The latter is dominated by two hadronic contributions: the hadronic vacuum polarization (HVP) and the hadronic light-by-light scattering (HLbL) contributions which respectively enter at order $\alpha^2$ and $\alpha^3$ where $\alpha$ is the electromagnetic coupling. 
The most precise determination of the HVP contribution, given by the dispersive approach, is a data driven estimates, and lattice QCD is the only rigorous tool to compute both contributions from first principle with reliable error estimates. 
In these proceedings, we summarize the status of our work~\cite{Gerardin:2019rua} in view of reducing the error below 0.5\%.

%---------------------------------------------------------------------------------------------------------
\section{Lattice setup and methodology}
%---------------------------------------------------------------------------------------------------------

This work is based on a set of ensembles from the Coordinated Lattice Simulations (CLS) initiative with $2+1$ dynamical quarks, listed in Table~\ref{tab:simul}, using the tree-level L\"uscher-Weiz gauge action and non-perturbatively O(a)-improved Wilson fermions. Four  lattice spacings in the range [0.050-0.086]~fm and several pion masses including the physical one are used to perform the extrapolation to the physical point. The charm quark is treated at the quenched level. More information about the ensembles can be found in Ref.~\cite{Bruno:2014jqa}. To further constrain the continuum extrapolation, the vector correlator is computed using two different discretizations of the vector current
\begin{eqnarray}
V_{\mu}^{\rm L}(x)  &=& \bar q(x) \gamma_{\mu}  q(x), \\
V_{\mu}^{\rm C}(x) &=& \frac{1}{2} \Big(\bar q(x + a\hat\mu ) ( 1 + \gamma_{\mu} ) U_{\mu}^{\dag}(x) q(x) - \bar q(x)( 1 - \gamma_{\mu} ) U_{\mu}(x) q(x + a\hat\mu ) \Big).
\end{eqnarray} 
The currents are non-perturbatively renormalized and O(a)-improved, as described in Ref.~\cite{Gerardin:2018kpy}. 
 
\begin{table}[t!]
        \caption{Parameters of the simulations. Ensembles E250 and B450 have periodic boundary conditions in time, all others have open boundary conditions. Ensembles with an asterisk are not included in the final analysis but are used to control finite-size effects.}        
\vskip 0.1in
\begin{centering}
{\footnotesize
\begin{tabular}{lcl@{\hskip 01em}c@{\hskip 01em}l@{\hskip 01em}l@{\hskip 01em}c@{\hskip 01em}c@{\hskip 01em}c@{\hskip 01em}c}
        \hline
        id     &       $\quad\beta\quad$       &       $L^3\times T$   &       $a\,[\fm]$       &    $~~~\kappa_l$          &       $~~~\kappa_s$      &
       $m_{\pi}\,[\MeV]$        &       $m_{K}\,[\MeV]$  &        $m_{\pi}L$     &       $L\,[\fm]$                     \\
        \hline
H101    & 3.40  &       $32^3\times96$  & 0.08636       &       0.136760        &       0.136760                & 416(5) & 416(5) & 5.8 & 2.8  \\  
H102    &               &       $32^3\times96$  &               &       0.136865        &       0.13654934      & 354(5) & 438(4) & 5.0 & 2.8  \\  
H105$^*$        &               &       $32^3\times96$  &               &       0.136970        &       0.13634079      & 284(4) & 460(4) & 3.9 & 2.8  \\         
N101    &               &       $48^3\times128$ &               &       0.136970        &       0.13634079      & 282(4) & 460(4) & 5.9 & 4.1  \\    
     
C101    &               &       $48^3\times96$  &               &       0.137030        &       0.13622204      & 221(2) & 472(8) & 4.7 & 4.1  \\  
\hline
B450            & 3.46  &       $32^3\times64$  & 0.07634       &       0.136890        &       0.136890                & 416(4) & 416(4) & 5.2 & 2.4  \\  
S400            &               &       $32^3\times128$  &               &       0.136984        &       0.13670239      & 351(4) & 438(5) & 4.3 & 2.4 \\          
N401    &               &       $48^3\times128$ &               &       0.137062        &       0.13654808      & 287(4) & 462(5)  & 5.3 & 3.7  \\  
\hline
H200$^*$        & 3.55  &       $32^3\times96$  & 0.06426       &       0.137000        &       0.137000                & 419(5) &  419(5) & 4.4 & 2.1  \\  
N202    &               &       $48^3\times128$ &               &       0.137000        &       0.137000                & 410(5) &  410(5) & 6.4 & 3.1  \\  
N203    &               &       $48^3\times128$ &               &       0.137080        &       0.13684028      & 345(4) &  441(5) & 5.4 & 3.1   \\  
N200    &               &       $48^3\times128$ &               &       0.137140        &       0.13672086      & 282(3) &  463(5) & 4.4 & 3.1   \\  
D200    &               &       $64^3\times128$ &               &       0.137200        &       0.13660175      & 200(2) &  480(5) & 4.2 & 4.1  \\  
E250    &               &       $96^3\times192$ &               &       0.137233        &       0.13653663      & 130(1) &    488(5)       & 4.1 & 6.2  \\  
\hline
N300    & 3.70  &       $48^3\times128$ &0.04981        &       0.137000        &       0.137000                & 421(4) & 421(4) & 5.1 & 2.4  \\  
N302    &               &       $48^3\times128$ &               &       0.137064        &       0.13687218      & 346(4) & 458(5) & 4.2 & 2.4  \\  
J303            &               &       $64^3\times192$ &               &       0.137123        &       0.13675466      & 257(3) & 476(5) & 4.2 & 3.2  \\  
\hline
 \end{tabular} }
\end{centering}
\label{tab:simul}
\end{table}

\begin{figure}
\floatbox[{\capbeside\thisfloatsetup{capbesideposition={right}}}]{figure}[\FBwidth]
{\caption{Integrand, at the physical pion mass, of the light, strange and charm quark contributions. For the strange and charm contribution, the integrand has been multiplied by a factor of six for clarity.}\label{fig:E250}}
{\includegraphics*[width=8.0cm]{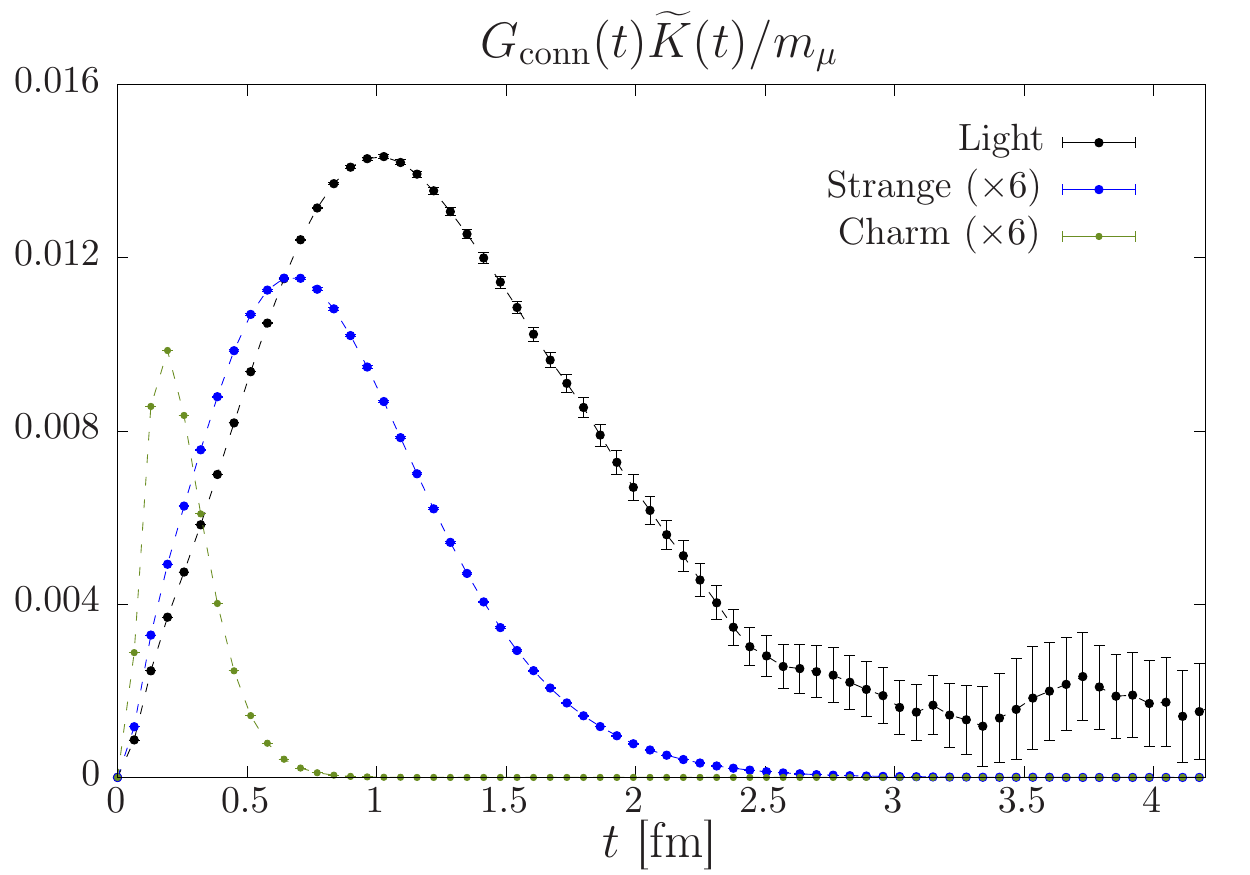}}
\end{figure}

In the time-momentum representation (TMR), the LO HVP contribution to the muon $(g-2)$ is expressed as a convolution integral between a known QED weight function $\widetilde K(t)$ and the vector two-point correlation function projected to vanishing momentum~\cite{Bernecker:2011gh}
\begin{equation}
\ahvp = \left(\frac{\alpha}{\pi}\right)^2\int_0^{\infty}\, dt\,\widetilde K(t)G(t), \quad G(t)\,\delta_{kl} = -\int d^3x \, \langle J_k(t,\vec x)\;J_l(0) \rangle \,.
\label{eq:master}
\end{equation}
In this work, we restrict ourselves to iso-symmetric QCD with $m_u=m_d$ and postpone the inclusion of isospin breaking corrections to future work. On the lattice, it is convenient to consider the flavor decomposition and to isolate the quark disconnected contribution
\begin{equation}
G(t) = \frac{5}{9} G_l(t) + \frac{1}{9}G_s(t) + \frac{4}{9} G_c(t) + G_{\rm disc}(t).
\end{equation}
The shape of the integrand for each connected quark contribution, and at the physical pion mass, is shown in Fig.~\ref{fig:E250}. The charm quark contribution is peaked at small distances and subject to large discretization effects. Full O(a)-improvement of the vector current is thus  necessary. For the light quark contribution, the signal-to-noise ratio deteriorates exponentially with time and advanced techniques must be used to reduce the noise. The light contribution is also affected by relatively large finite-size effects (FSE) that must be properly accounted for. Finally, quark disconnected contributions, notoriously difficult to estimate in lattice QCD, contribute to about $-2$\% of the total contribution. In the next sections, we explain how we address all these issues.
 
\vspace{-1.1cm}

%---------------------------------------------------------------------------------------------------------
\section{Results}
%---------------------------------------------------------------------------------------------------------

%---------------------------------------------------------------------------------------------------------
\subsection{Solution to the noise problem}
%---------------------------------------------------------------------------------------------------------

\begin{figure}[h]
	\centering
	\includegraphics*[width=0.45\linewidth]{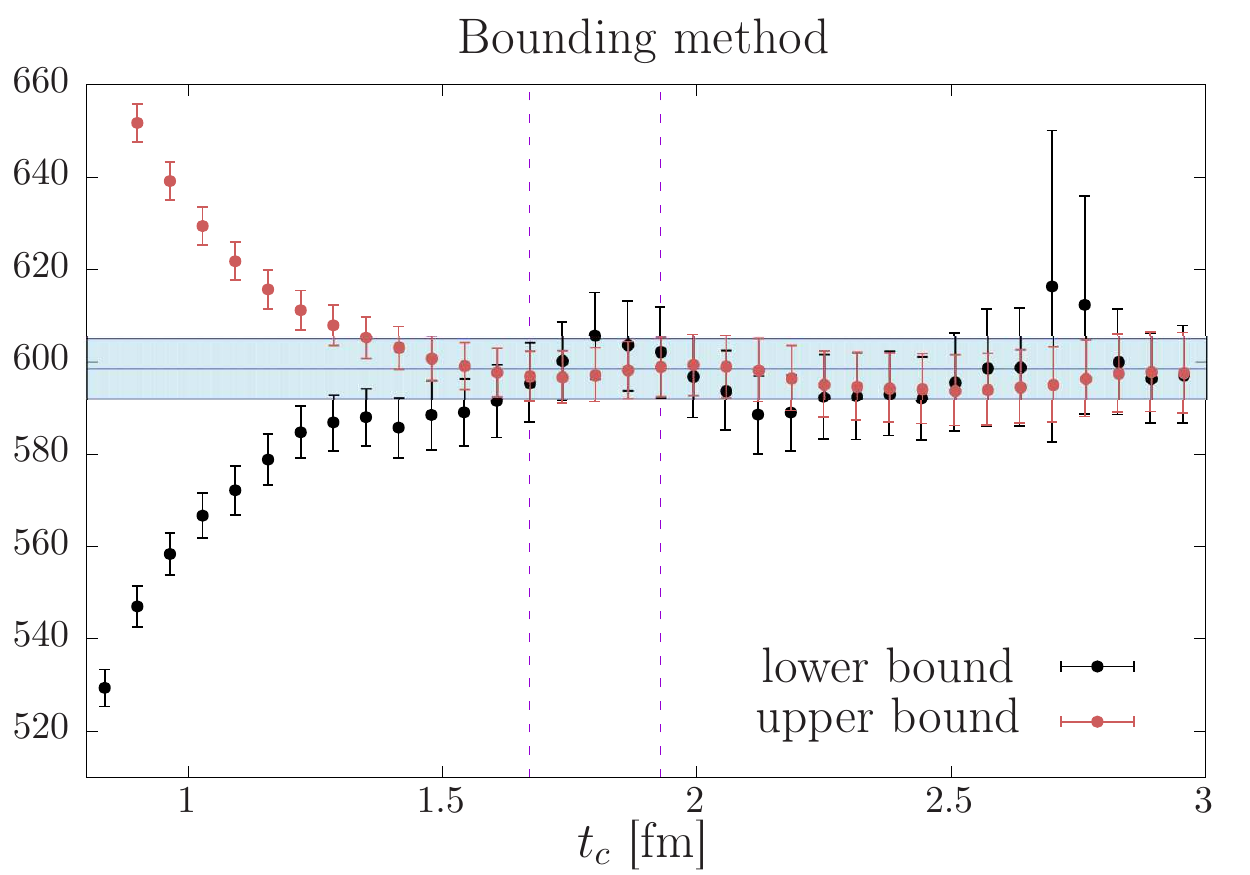}
	\includegraphics*[width=0.45\linewidth]{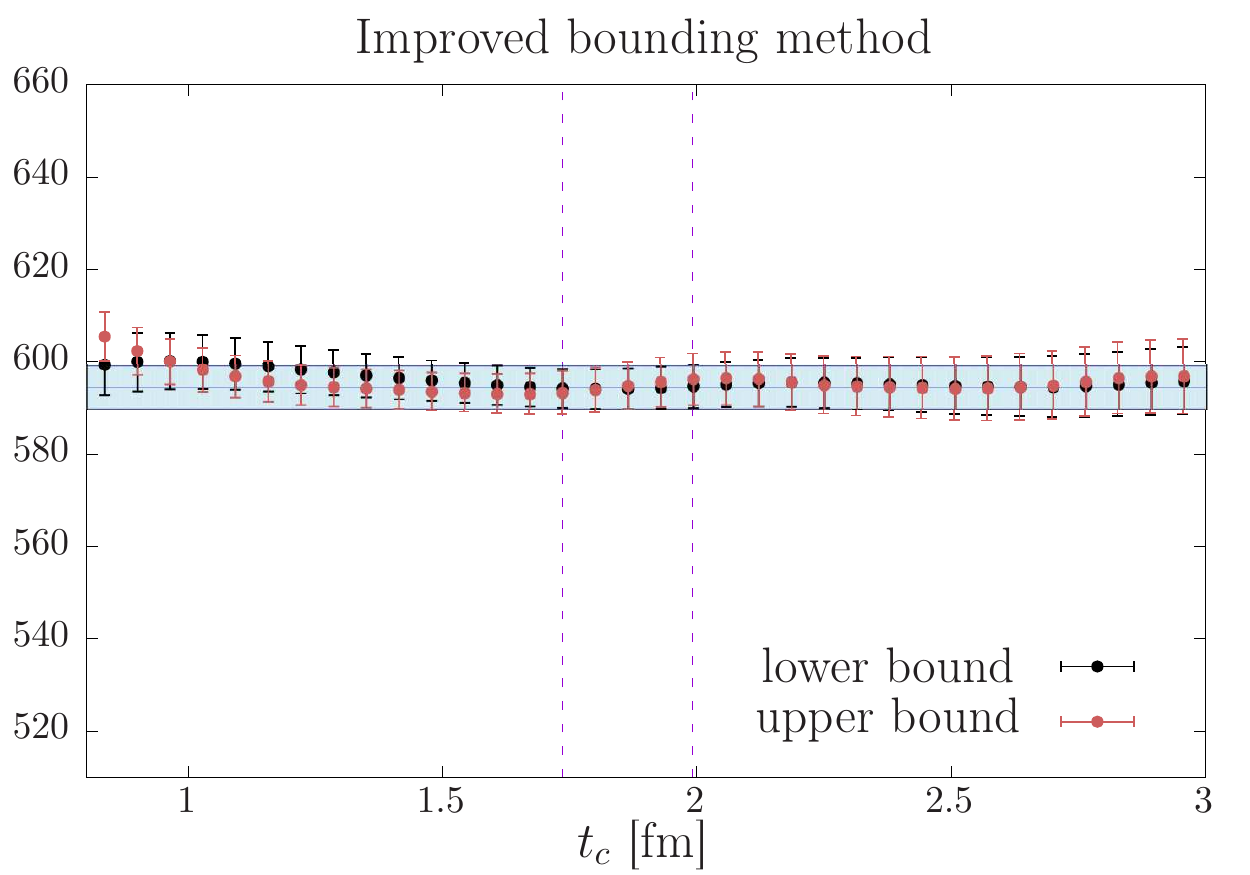}
			
	\caption{Bounding (left) and improved bounding (right) methods applied to the ensemble D200. The final result is obtained by averaging the data between the two vertical dashed lines.}
	\label{fig:bounding}
\end{figure}

\begin{figure}[h]
	\centering
	\includegraphics*[width=0.45\linewidth]{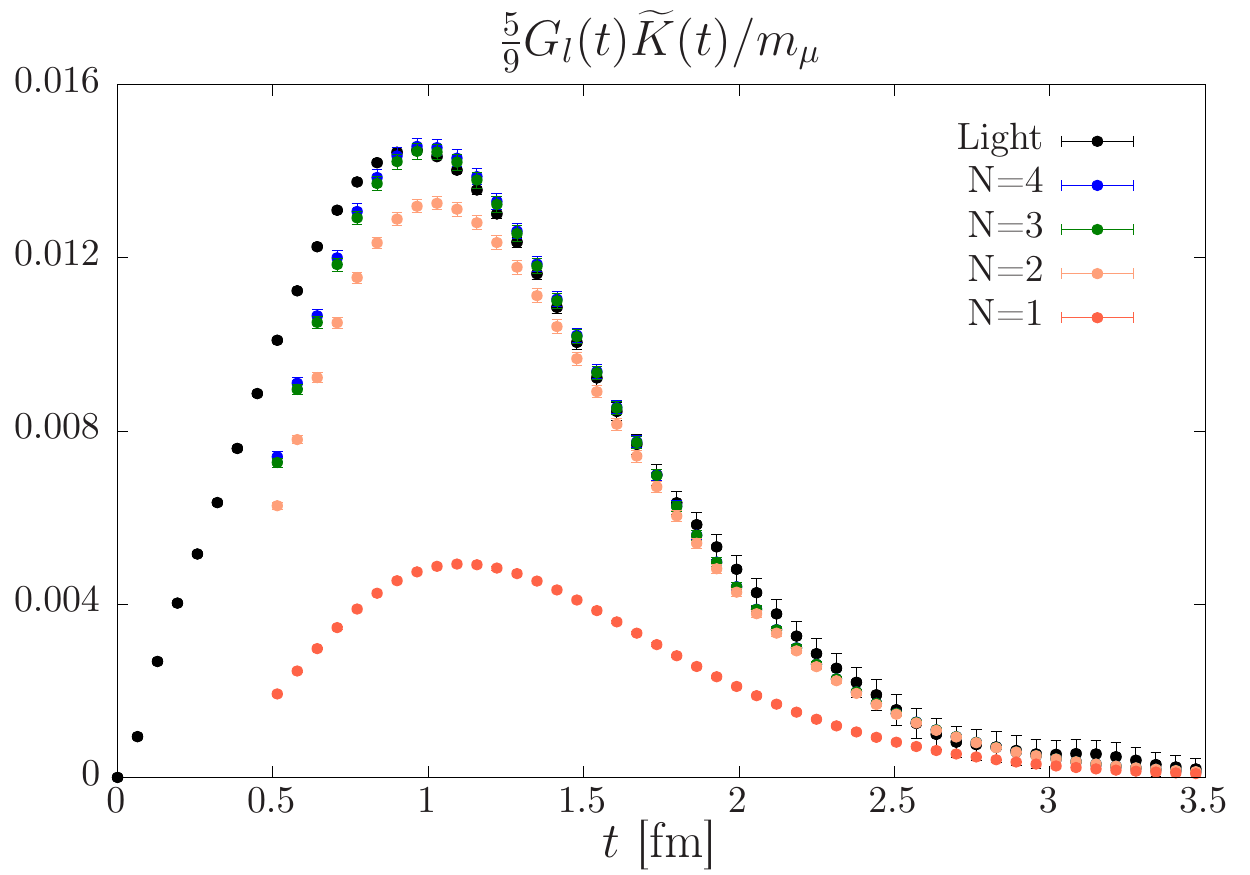}
	\includegraphics*[width=0.45\linewidth]{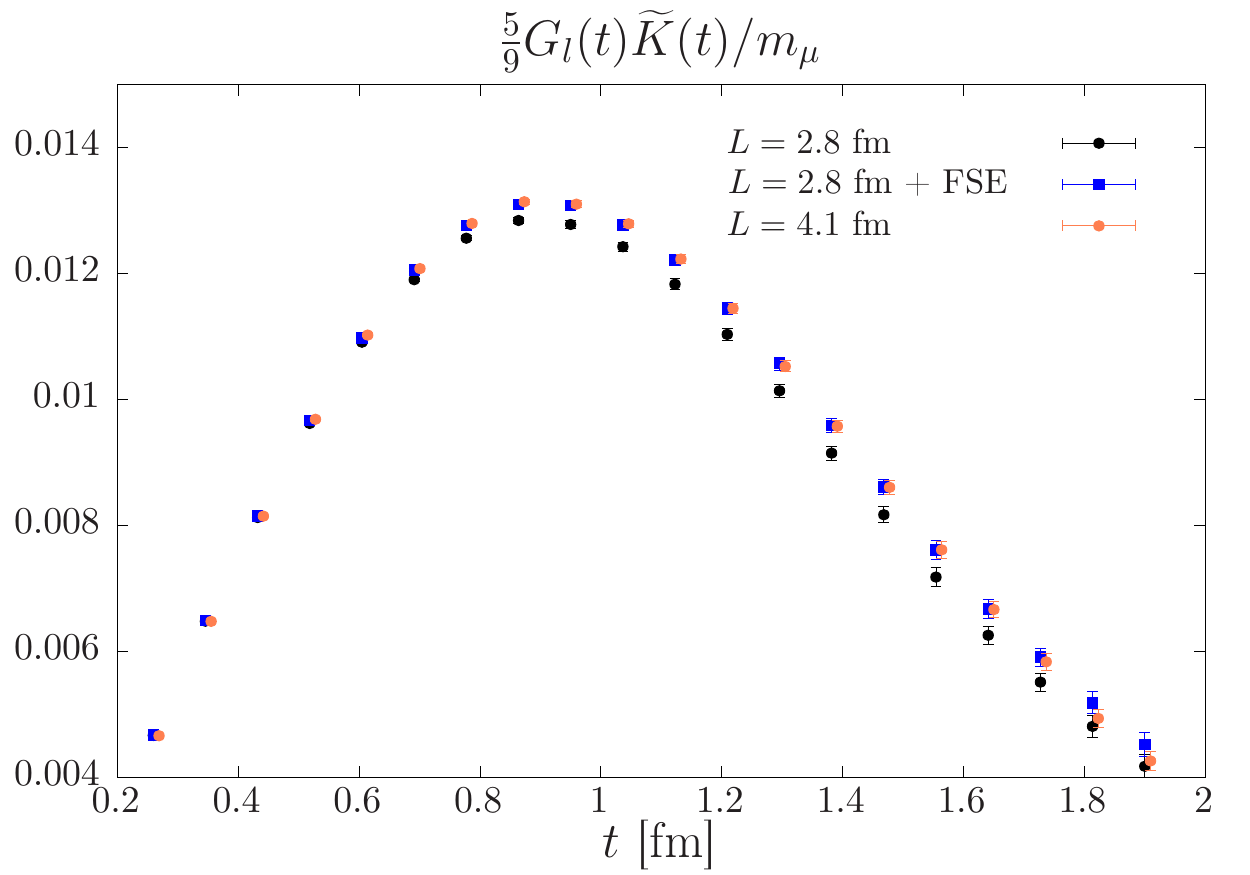}
		
	\caption{Left: reconstruction of the light correlator using Eq.~(\ref{eq:specdec}) for different values of $n$. Right: check of the FSE correction method described in Sec.~\ref{sec:FSE}. After correction of the lattice data obtained in a small volume (blue points), the black points agree perfectly with the large volume simulation, in orange.}	
	\label{fig:syst}
\end{figure}

One of the main challenge to improve the statistical precision is to control the exponential growth of the noise-over-signal ratio at large time for the light contribution (see Fig.~\ref{fig:E250}). The bounding method~\cite{Borsanyi:2017zdw} provides a systematic way to cut the integration range in Eq.~(\ref{eq:master}). It relies on the observation that the correlator satisfies the two rigorous bounds
\begin{equation}
0\leq G(t_c) e^{-E_{\rm eff}(t_c)(t-t_c)} \leq G(t) \leq G(t_c) e^{-E_N(t-t_c)}, \qquad t\geq t_c,
\end{equation}
where $t_c$ can be chosen such that both bounds agree within statistical precision (left panel of Fig~\ref{fig:bounding}). 
This method can be improved by using the spectral decomposition of the vector correlator 
\begin{equation}
G(t) = \sum_{n=0}^\infty \, \frac{Z_n^2}{2E_n} \, e^{-E_n t}.
\label{eq:specdec}
\end{equation}
At a given statistical precision, and sufficiently large times, only a small number of states ($N$) is needed to saturate the sum. Using a large basis of interpolating operators and distillation techniques~\cite{Andersen:2018mau}, we are able to extract the energies and overlaps of the first low-lying states. The shape of the integrand using the reconstruction of the correlator for various values of $N$ is shown in Fig.~\ref{fig:syst}. The statistical error on the truncated correlator now grows linearly with time, solving the noise problem. The improved bounding method consists in applying the bounding method to the subtracted correlator
\begin{equation}
\widetilde G(t) = G(t) - \sum_{n=0}^{N-1} \frac{Z_n^2}{2E_n} e^{-E_n t}.
\end{equation}
The results for a pion mass of $200$~MeV are shown in Fig.~\ref{fig:bounding} where we are able to reach a statistical precision of the order of 0.7\%. 

%---------------------------------------------------------------------------------------------------------
\subsection{Correction for finite-size effects}
\label{sec:FSE}
%---------------------------------------------------------------------------------------------------------

For the ensemble at the physical pion mass, with $m_{\pi} L \geq 4$, we estimate the finite-size effect correction to be of the order of 3\%. Since lattice calculations aim at an overall precision below 0.5\% in the next few years, it is of major importance to treat them carefully. Chiral perturbation theory is expected to work only at asymptotically large volumes and we prefer to use the method described in Ref.~\cite{Bernecker:2011gh}. The isovector correlator in infinite and finite volume reads
\begin{eqnarray}
G^{I=1}(t, \infty) &=& \int_{2m_{\pi}}^{\infty} \mathrm{d}\omega \, \omega^2 \, \rho(\omega^2)\, e^{-\omega t}  \,, \quad \rho(\omega^2) = \frac{1}{48\pi^2} \left(1 - \frac{4 m_{\pi}^2}{\omega^2} \right)^{3/2}  |F_{\pi}(\omega)|^2  \\
G^{I=1}(t, L) &=& \sum_{i} |A_i|^2 \, e^{- E_i t} \,, \quad  E_i = 2 \sqrt{ m_{\pi}^2 +  k_i^2}  \,,
\end{eqnarray}
where $F_{\pi}$ is the timelike pion form factor. The difference between these two formulae leads to our FSE estimate. Using the L\"uscher formalism~\cite{Luscher:1991cf,Meyer:2011um}, the finite volume energies $E_i$ and the overlap factors $|A_i|^2$ are related to the 
scattering phase shift in the isospin $I=1$, p-wave channel and to $F_{\pi}$~\cite{Meyer:2011um}. The key ingredient is therefore $F_{\pi}$, which has been computed on a subset of our ensembles~\cite{Andersen:2018mau}. We have explicitly checked the validity of this approach using two lattice simulations performed at $m_{\pi} \approx 280$~MeV and with volumes of $2.8$ and $4.1$~fm respectively. The results are depicted on the right panel of Fig.~\ref{fig:syst} where the corrected lattice data, obtained in a volume with $m_{\pi} L \approx 4$, perfectly agree with the large volume simulation.

%---------------------------------------------------------------------------------------------------------
\subsection{The quark-disconnected contribution}
%---------------------------------------------------------------------------------------------------------

\begin{figure}[t]
	\centering
	\vspace{-0.45cm}	
	\includegraphics*[width=0.45\linewidth]{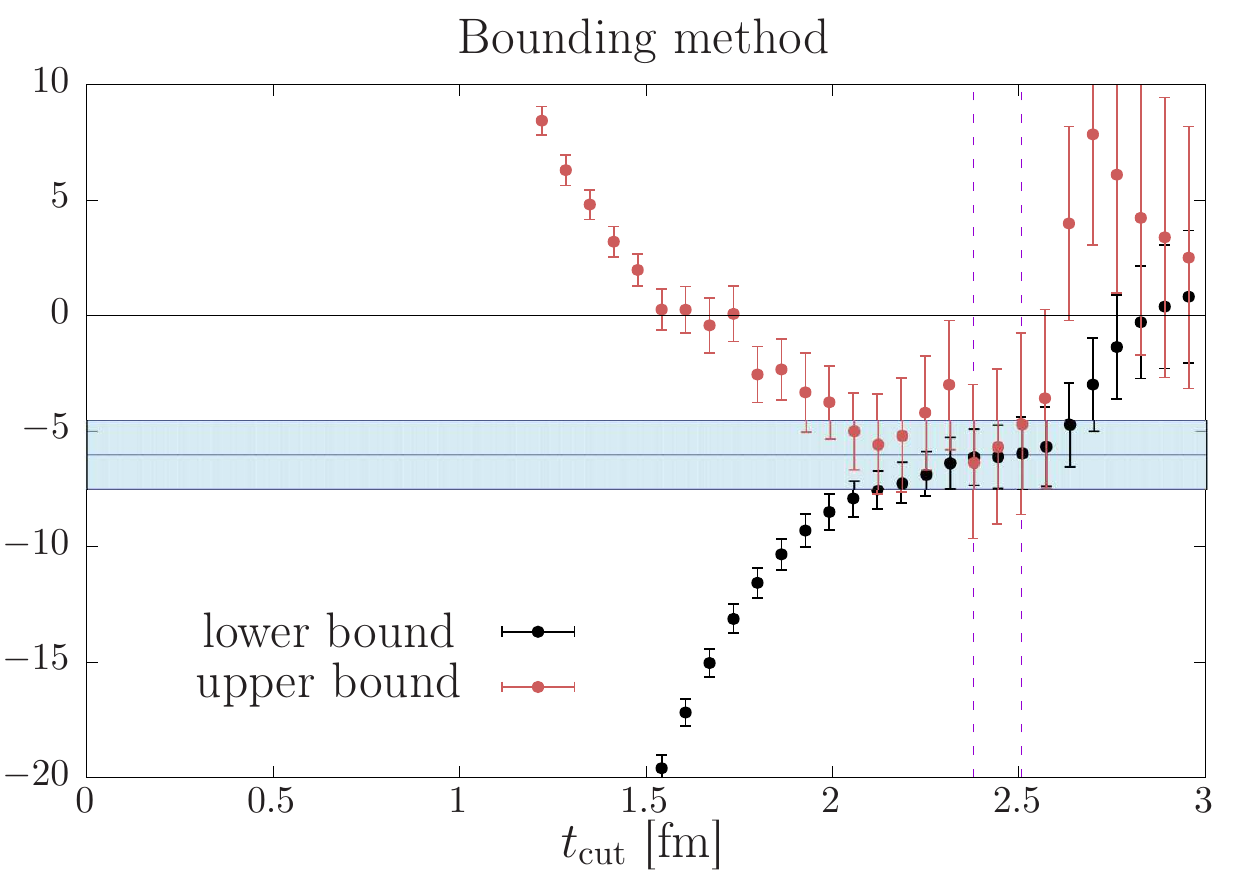}
	\includegraphics*[width=0.45\linewidth]{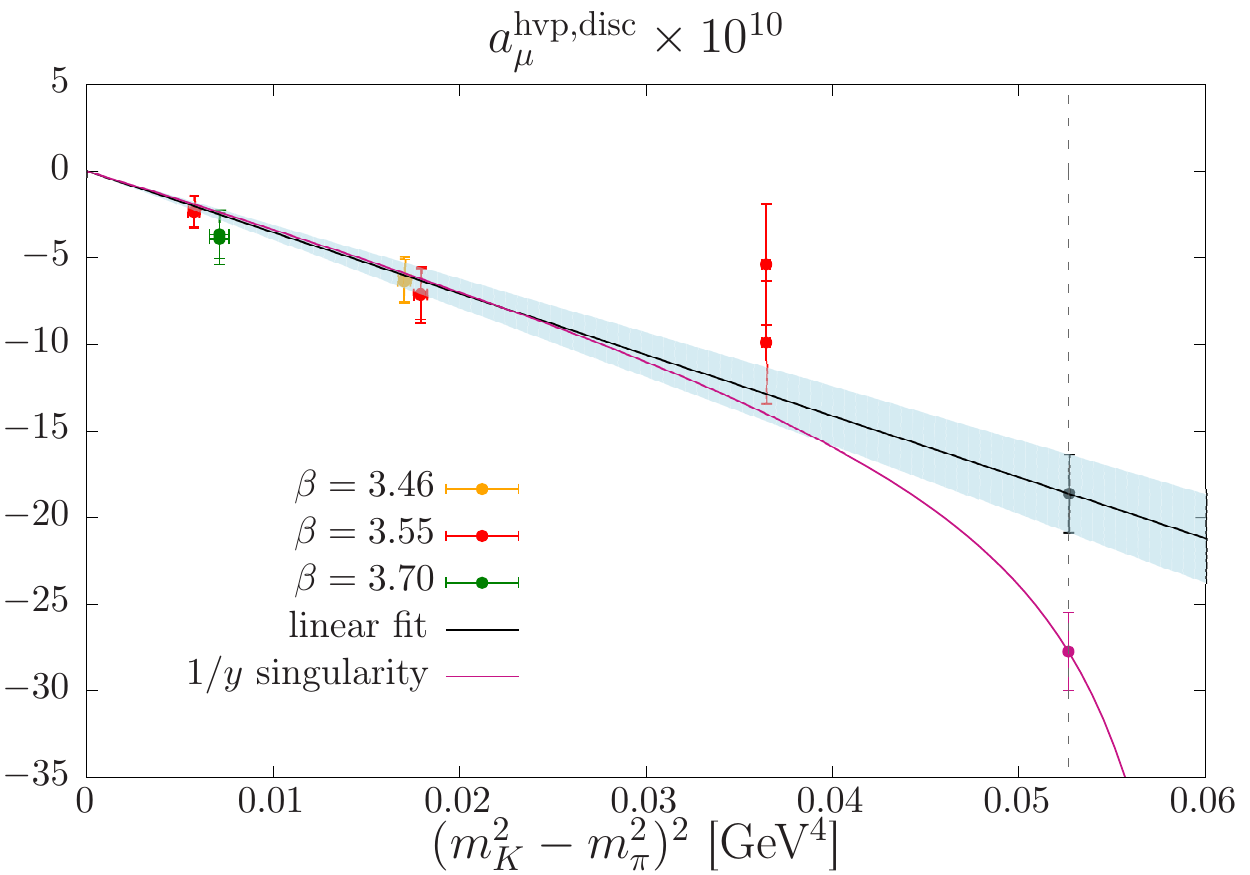}
	
	\vspace{-0.05cm}		
	\caption{Left: bounding method applied to the isoscalar correlator (the light and strange quark contributions have been subtracted). Right: extrapolation of the quark disconnected contribution to the physical point.}	
	\label{fig:disc}
\end{figure}

The quark-disconnected contribution has been computed on a subset of ensembles using noise reduction techniques~\cite{Djukanovic:2018iir}. The signal deteriorates rapidly at large time and is lost at $t\approx 1.5~$fm. To constrain the tail of this contribution, we apply the bounding method to the isoscalar correlator
\begin{equation}
G(t) = G^{I=1}(t) + G^{I=0}(t),\qquad 
G^{I=1}(t) = \frac{1}{2} G_l(t),
\end{equation}
\begin{equation}
0\leq G^{I=0,\slashed{c}}(t) \leq G^{I=0,\slashed{c}}(t_c) e^{-m_\rho(t-t_c)}, \qquad t\geq t_c.
\end{equation}
Here, $\slashed{c}$ means that the charm contribution is not included. The disconnected contribution is obtained by subtracting the precisely known light and strange contributions and the data are corrected for finite size effects. Results for the ensemble N200 are shown in Fig.~\ref{fig:disc}. At our level of precision, we do not observe significant discretization effects and the extrapolation to the physical point is performed assuming the ansatz (with a single fit parameter $\gamma_8$)
\begin{equation}
a_\mu^{{\rm hvp,\,disc}}(\Delta_2) = \gamma_8 \Delta_2^{\,2} -\frac{\alpha^2m_\mu^2}{3240\pi^2} \cdot
\frac{3}{2}\Big[ \frac{1}{\hat M^2 - \Delta_2}   - \frac{\Delta_2}{\hat M^4} - \frac{1}{\hat M^2}\Big],
\end{equation}
where $\hat M^2 \equiv \frac{1}{2}m_\pi^2 + m_K^2$ is almost constant along our chiral trajectory and $\Delta_2\equiv m_K^2-m_\pi^2$. 
The singular contribution $\propto m_{\mu}^2/m_{\pi}^2$ is designed to cancel against the connected part in the isoscalar contribution. The result of the chiral extrapolation, shown on the right panel of Fig.~\ref{fig:disc}, reads
\begin{equation}
a_\mu^{{\rm hvp,\,disc}}= (-23.2 \pm 2.2 \pm 4.5)\times 10^{-10},
\end{equation}
where the first error is statistical and the second error accounts for the chiral extrapolation. Clearly,~more data close to and at the physical pion mass are needed to remove this large systematic uncertainty.

%---------------------------------------------------------------------------------------------------------
\subsection{Strange and charm quark contributions}
%---------------------------------------------------------------------------------------------------------

\begin{figure}[h]
	\centering
	\includegraphics*[width=0.45\linewidth]{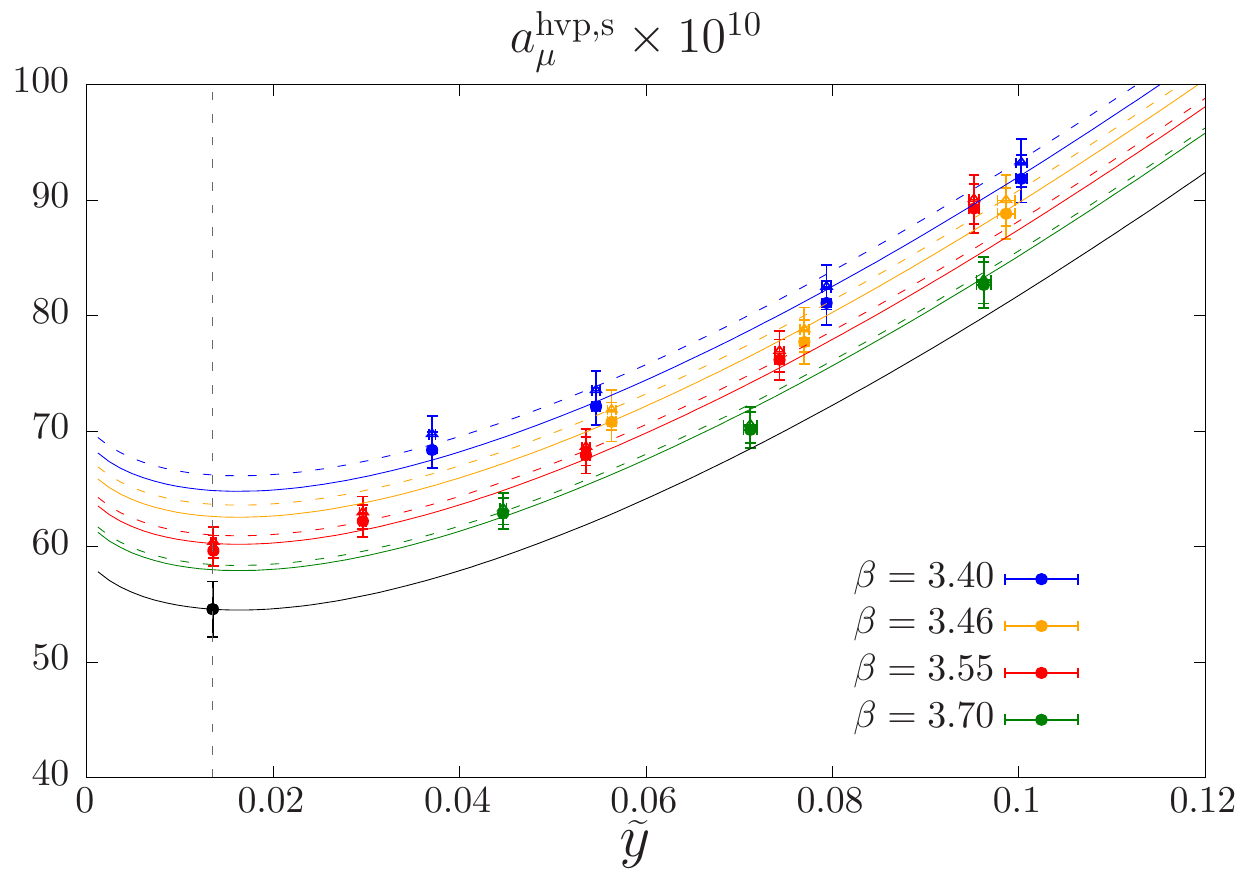}
	\includegraphics*[width=0.45\linewidth]{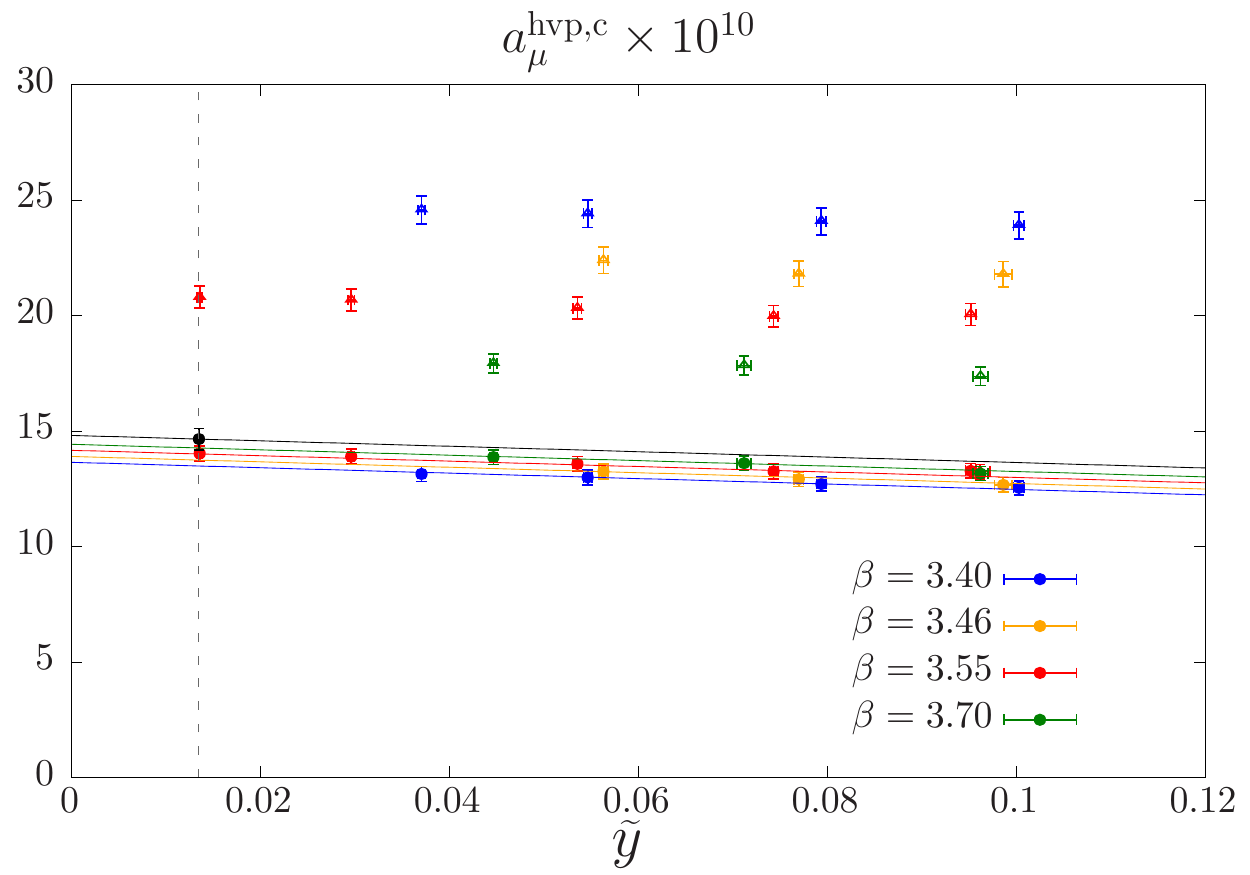}
			
	\vspace{-0.05cm}
	\caption{Continuum and chiral extrapolation of the strange (left) and charm (right) quark contribution.}	
	\label{fig:cs}
\end{figure}

For the strange and charm quarks, FSE are negligible and no sophisticated treatment of the tail is needed: the bounding method is used. In Fig.~\ref{fig:cs} we show the extrapolation to the physical point for both contributions. Note that the charm quark is quenched and that we do not use the local-local discretization of the correlator which is affected by large discretization effects. The results are
\begin{equation}
a_\mu^{{\rm hvp},s} = (54.5\pm 2.4\pm 0.6)\times 10^{-10} \,,\quad
a_\mu^{{\rm hvp},c} = (14.66\pm 0.45 \pm 0.06)\times 10^{-10}\,,
\end{equation}
where the first error is statistical and the second is the systematic error from the chiral extrapolation. 
The fraction of the strange and charm quark uncertainties over the total value are 0.34\% and 0.06\% respectively. 
In both cases, the error is dominated by the scale setting uncertainty such that it can be significantly reduced once an improved scale setting is performed.

\vspace{-0.4cm}
%---------------------------------------------------------------------------------------------------------
\section{Conclusion}
%---------------------------------------------------------------------------------------------------------

\begin{figure}[t]
	\centering
	\vspace{-0.3cm}	
	\includegraphics*[width=0.7\linewidth]{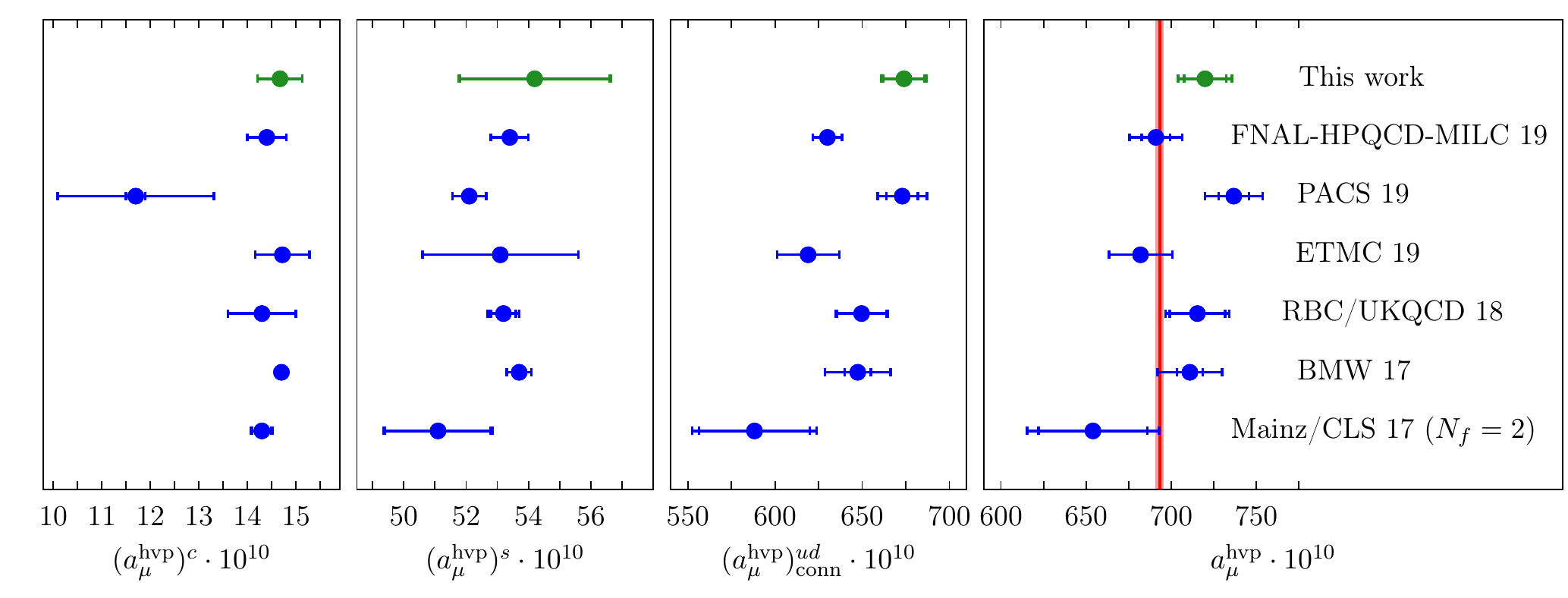}
			
	\vspace{-0.19cm}	
	\caption{Comparison with other recent lattice calculations~\cite{Borsanyi:2017zdw,Giusti:2018mdh,Blum:2018mom,Davies:2019efs,Shintani:2019wai}.}
	\label{fig:cmp}
\end{figure}

We have presented a calculation of the LO HVP contribution to the anomalous magnetic moment of the muon using Wilson quarks. Our  result, at the physical point, reads $\ahvp = (720.0\pm 12.4 \pm 6.8)\times 10^{-10}$ and corresponds to a relative precision of 2\%. A comparison with recent lattice calculations is given in Fig.~\ref{fig:cmp}. The error is dominated by statistics. We have shown that a precise knowledge of the spectrum in the vector channel allows us to reduce significantly the statistical error in the light contribution and we plan to apply this procedure directly at the physical~pion mass. Another ensemble, at the finest lattice spacing and with $m_{\pi} \approx 175~$MeV will be added. More statistics on the disconnected contribution is also underway and we plan to include isospin breaking effects from a dedicated lattice simulation~\cite{Risch:2018ozp}. All these improvements, together with a better scale-setting determination, should allow us to reach a precision below 0.5\% in the near future.

\vspace{0.1cm}
{\small\textbf{Acknowledgments :}
This work is partly supported by the DFG grant HI 2048/1-1
and by the DFG-funded Collaborative Research Centre SFB\,1044 \emph{The low-energy frontier of the Standard Model}.
The Mainz $(g-2)_\mu$ project is also supported by the Cluster of Excellence \emph{Precision Physics, Fundamental Interactions, and Structure of Matter} (PRISMA+ EXC 2118/1) funded by the DFG within the German Excellence Strategy (Project ID 39083149).
Calculations for this project were partly performed on the HPC clusters ``Clover'' and ``HIMster II'' at the Helmholtz-Institut Mainz and ``Mogon II'' at JGU Mainz.  }

\vspace{-0.30cm}	
%---------------------------------------------------------------------------------------------------------

\end{document}